\documentclass[aps,prd,12pt,notitlepage,nofootinbib,tightenlines]{revtex4-1}
\usepackage{amsmath}
\usepackage{bm}
\usepackage{times}
\usepackage{braket}
\usepackage{color,graphicx,epsfig}
\usepackage{slashed}
\usepackage{hyperref}
\newcommand{\beq}{\begin{eqnarray}}
\newcommand{\eeq}{\end{eqnarray}}
\newcommand{\non}{\nonumber\\ }
\newcommand{\ov}{\overline}

\newcommand{\mw}{m_W }

\newcommand{\kt}{k_{\rm T} }

\newcommand{\etap}{ \eta^{(')} }

\newcommand{\acp}{{\cal A}_{\rm CP}}

\newcommand{\psl}{ P \hspace{-2.4truemm}/ }

\newcommand{\calm}{ {\cal M} }

\definecolor{Red}{rgb}{1.,0.,0.}

\definecolor{Blue}{rgb}{0.,0.,1.}

\definecolor{nicered}{rgb}{0.7,0.1,0.1}
\definecolor{nicegreen}{rgb}{0.1,0.5,0.1}

\hypersetup{colorlinks,citecolor=nicegreen,linkcolor=nicered}

\begin{document}

\def \cpc{ Chin. Phys. C }
\def \epjc{ Eur. Phys. J. C }
\def \jpg{  J. Phys. G }
\def \npb{  Nucl. Phys. B }
\def \plb{  Phys. Lett. B }
\def \pr{  Phys. Rep. }
\def \prd{  Phys. Rev. D }
\def \prl{  Phys. Rev. Lett.  }
\def \zpc{  Z. Phys. C  }
\def \jhep{ J. High Energy Phys.  }

\title{$B\to \pi\pi$ decays and effects of the next-to-leading order contributions}
\author{Ya-Lan Zhang$^1$,  Xue-Yan Liu$^1$, Ying-Ying Fan$^1$, Shan Cheng$^1$,and Zhen-Jun Xiao$^{1,2}$
\footnote{Email Address: xiaozhenjun@njnu.edu.cn} }
\affiliation{1. Department of Physics and Institute of Theoretical Physics,\\
Nanjing Normal University, Nanjing, Jiangsu 210023, P.R. China}
\affiliation{2. Jiangsu Key Laboratory for Numerical Simulation of Large Scale Complex Systems,
Nanjing Normal University, Nanjing 210023, People's Republic of China}
\date{\today}
\begin{abstract}
In this paper we perform a systematic study for the three $B \to (\pi^+\pi^-,\pi^+\pi^0,\pi^0\pi^0)$
decays in the perturbative QCD (pQCD) factorization approach with the inclusion of all currently known
next-to-leading order (NLO) contributions from various sources. We found that
(a) for the CP-averaged decay rates $Br(B^0\to \pi^+\pi^-)$ and $Br(B^+\to \pi^+\pi^0)$,
the NLO pQCD predictions agree with the data within one standard dviation;
(b) for $Br(B^0\to \pi^0\pi^0)$, however, although the NLO contributions can provide
a $\sim 100\%$ enhancement to the leading order (LO) result, it is still not large enough to interpret the data;
(c) for the CP-violating asymmetries of $B^0\to \pi^+\pi^-$ decay, the central values of the
NLO PQCD predictions agree with the data; and
(d) we also examined the relative strength of the LO and NLO contributions from different sources.
\end{abstract}


\maketitle

As is well-known, the standard model (SM) prediction for $Br(B^0\to\pi^0\pi^0)$ \cite{lu2001,by05,nlo05}
is much smaller than the measured one, which has been known as the ``$\pi\pi$"
puzzle in $B\to \pi\pi$ decays \cite{hfag2012,pdg2012}.
In Ref.~\cite{nlo05}, the authors studied this puzzle by employing
the PQCD approach \cite{li92,huang91,li2003}
by including partial next-to-leading order (NLO) contributions known at that time, and
found that $Br(B^0\to\pi^0\pi^0)$ can be increased from the leading order (LO) prediction $0.12\times 10^{-6}$
to $0.29\times 10^{-6}$.

In Refs.~\cite{prd85-074004,cheng14a,cheng14b}, very recently, the authors calculated the
NLO twist-2 and twist-3 contributions to the form factors of $B \to \pi$ transition in the pQCD approach.
We here will study the $B \to \pi\pi$ decays again with the inclusion of these newly known NLO contributions
to form factors and to check their effects.

In the B-rest frame,  we assume that the light final state pion mesons
are moving along the direction of $n=(1,0,{\bf{0}}_T)$ and $v=(0,1,{\bf{0}}_T)$, respectively.
We use $x_i$ to denote the momentum fraction of the anti-quark in each meson,
$\kt$ the corresponding transverse momentum. Using the light-cone coordinates
the $B$ meson momentum $P_B$ and the two final state pion meson's momenta $P_2$ and $P_3$ can be written as
\beq
P_B = \frac{M_B}{\sqrt{2}} (1,1,{\bf 0}_{\rm T}), \quad
P_2 =\frac{M_B}{\sqrt{2}}(1-r_3^2,r^2_2,{\bf 0}_{\rm T}), \quad
P_3 =\frac{M_B}{\sqrt{2}} (r_3^2,1-r^2_2,{\bf 0}_{\rm T}),
\eeq
where $r_i=m_\pi/M_B$. After the integration over the small components $k_1^-$, $k_2^-$, and $k_3^+$  we find
the decay amplitudes conceptually
\beq
{\cal A}(B \to M_2M_3 ) &\sim &\int\!\! d x_1 d x_2 d x_3 b_1 d b_1 b_2 d b_2 b_3 d b_3 \non &&
\cdot \mathrm{Tr} \left [ C(t) \Phi_B(x_1,b_1) \Phi_\pi (x_2,b_2)
\Phi_\pi(x_3, b_3) H(x_i, b_i, t) S_t(x_i)\, e^{-S(t)} \right ],
\quad \label{eq:a2}
\eeq
where $b_i$ is the conjugate space coordinate of $k_{iT}$, $C(t)$ is the Wilson
coefficient, the functions $\Phi_B(x_1,b_1)$, $\Phi_\pi(x_j,b_j)$ with $j=(2,3)$
are the wave functions of the initial B meson and the two
final state pion mesons respectively. The function $H(k_1,k_2,k_3,t)$ is the hard kernel, while
the jet function $S_t(x_i)$ and the function $e^{-S(t)}$ are the two Sudakov factors relevant
for the considered B decays \cite{li2003}.

For the considered  $B \to \pi\pi $ decays, the corresponding weak
effective Hamiltonian can be written as \cite{buras96}:
\beq
{\cal H}_{eff} &=& \frac{G_{F}}{\sqrt{2}}     \Bigg\{ V_{ub} V_{ud}^{\ast} \Big[
     C_{1}({\mu}) O^{u}_{1}({\mu})  +  C_{2}({\mu}) O^{u}_{2}({\mu})\Big]
  -V_{tb} V_{td}^{\ast} \Big[{\sum\limits_{i=3}^{10}} C_{i}({\mu}) O_{i}({\mu})
  \Big ] \Bigg\} + \mbox{H.c.} ,  \label{eq:heff}
\eeq
where $G_{F}=1.166 39\times 10^{-5} GeV^{-2}$ is the Fermi constant,
$V_{ij}$ are the elements of the
Cabbibo-Kobayashi-Maskawa (CKM) quark mixing matrix, the $O_{i}$
($i=1,...,10$) are the local four-quark operators and $C_i(\mu)$ are the Wilson coefficients
evaluated at scale $\mu$ \cite{buras96}.

The $B$ meson is treated as a very good heavy-light system with
the wave function in the form of
\beq
\Phi_B= \frac{i}{\sqrt{2N_c}} (\psl_B +m_B) \gamma_5 \phi_B ({\bf k_1}).
\label{bmeson}
\eeq
Here we adopted the B-meson distribution amplitude  $\phi_B(x,b)$ widely used
for example in Refs.~\cite{lu2001,prd65-014007}
\beq
\phi_B(x,b)&=& N_B x^2(1-x)^2 \exp \left  [ -\frac{M_B^2\ x^2}{2 \omega_{b}^2} -\frac{1}{2} (\omega_{b} b)^2\right],
 \label{phib}
\eeq
where the $b$-dependence was included through the second term in the exponential
function, the shape parameter $\omega_b =0.40\pm 0.04$ has been fixed \cite{li2003}
from the fit to the $B \to \pi$ form factors derived from lattice QCD and from Light-cone sum
rule \cite{ball98}, and finally the normalization factor $N_B$ depends on the value of
$\omega_b$ and $f_B$ and defined through the normalization relation:
$\int_0^1dx \; \phi_B(x,b=0)=f_B/(2\sqrt{6})$. The wave functions
of the final state pion mesons
and the relevant distribution amplitudes $\phi_\pi^{A,P,T}$ are of the same form as being
adopted in Refs.~\cite{nlo05,xiao2008,fan2013,xiao2013}. The Gegenbauer moments $a_i^\pi$ and
other parameters are adopted from Refs.~\cite{ball05,nlo05}:
\beq
a_2^{\pi} &=& 0.25, \quad a_4^{\pi} = -0.015, \quad \rho_{\pi} = m_{\pi}/m_{0\pi}, \quad \eta_3 = 0.015,
\quad \omega_3 = -3.0,
\label{eq:input1}
\eeq
with $m_{0\pi}$ is the chiral mass of pion.

\begin{figure}[tb]
\vspace{-6cm}
\centerline{\epsfxsize=16cm \epsffile{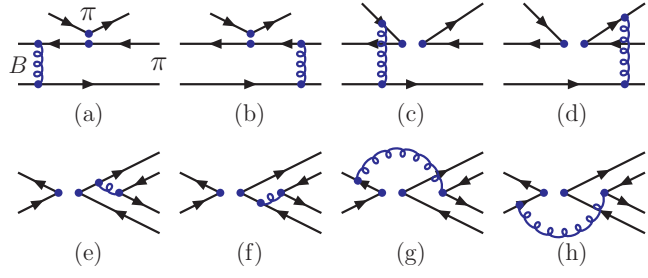}}
\vspace{-14cm}
\caption{ Feynman diagrams which may contribute to the $B\to \pi\pi$ decays in the pQCD approach at leading order.}
\label{fig:fig1}
\end{figure}

The $B \to \pi\pi$ decays have been studied by employing the pQCD factorization
approach at the LO ~\cite{lu2001} or partial NLO level \cite{nlo05}. The total decay amplitude
at the leading order for the three  $B \to \pi\pi$ decays are the following
\beq
\calm_{\rm LO}(B^0\to \pi^+\pi^-) &=& \frac{G_F}{\sqrt{2}}
\Biggl\{\lambda_u\Big[a_1f_{\pi}F_{e\pi}^{V-A}+c_1 M_{e\pi}^{V-A}
 +a_2f_B F_{a\pi}^{V-A}+c_2M_{a\pi}^{V-A}\Big]\nonumber\\
&& \hspace{-2.5cm}-\lambda_t\Big[\left (a_4+a_{10}\right )f_{\pi }F_{e\pi}^{V-A}
 +\left (a_6+a_8\right )f_{\pi } F_{e\pi}^{SP}
 +\left (c_3+c_9\right )M_{e\pi}^{V-A}
+\left (c_5+c_7\right )M_{e\pi}^{V+A} \non
&& \hspace{-2.5cm}
+\left (2a_3+a_4+\frac{1}{2}a_9-\frac{1}{2}a_{10}\right )f_B F_{a\pi}^{V-A}
+\left(2a_5+\frac{1}{2}a_7\right)f_B F_{a\pi}^{V+A}
 +\left(a_6-\frac{1}{2}a_8\right)f_B F_{a\pi}^{SP} \non
&& \hspace{-2.5cm}
+\left(c_3+2c_4-\frac{1}{2}c_9+\frac{1}{2}c_{10}\right)M_{a\pi}^{V-A}
+\left(c_5-\frac{1}{2}c_7\right)M_{a\pi}^{V+A}
+\left(2c_6+\frac{1}{2}c_8\right)M_{a\pi}^{SP}\Big]\Biggr \}, \label{eq:pipi1}
\eeq
\beq
\calm_{\rm LO}(B^0\to \pi^0\pi^0)&=&\frac{1}{\sqrt{2}}\frac{G_F}{\sqrt{2}}
\Biggl\{\lambda_u\Bigl[-a_2f_{\pi}F_{e\pi}^{V-A}-c_2M_{e\pi}^{V-A}+a_2f_B F_{a\pi}^{V-A}+c_2M_{a\pi}^{V-A}\Bigr]\non
&&\hspace{-2cm}-\lambda_t\Bigl[ (-\frac{3}{2}a_7)f_{\pi}F_{e\pi}^{V+A}
+\left(a_4-\frac{3}{2}a_9-\frac{1}{2}a_{10}\right)f_\pi F_{e\pi}^{V-A}
+\left(a_6-\frac{1}{2}a_8\right)f_{\pi}F_{e\pi}^{SP}\non
&&\hspace{-2cm}+\left(c_3-\frac{1}{2}c_9-\frac{3}{2}c_{10}\right)M_{e\pi}^{V-A}
+\left(c_5-\frac{1}{2}c_7\right)M_{e\pi}^{V+A}
+\left(-\frac{3}{2}c_8\right)M_{e\pi}^{SP}\non
&&\hspace{-2cm}+\left(2a_3+a_4+\frac{1}{2}a_9-\frac{1}{2}a_{10}\right)f_B F_{a\pi}^{V-A}
+\left(2a_5+\frac{1}{2}a_7\right)f_B F_{a\pi}^{V+A}
+\left(a_6-\frac{1}{2}a_8\right)f_B F_{a\pi}^{SP}\non
&&\hspace{-2cm}+\left(c_3+2c_4-\frac{1}{2}c_9+\frac{1}{2}c_{10}\right)M_{a\pi}^{V-A}
+\left(c_5-\frac{1}{2}c_7\right)M_{a\pi}^{V+A}
+\left(2c_6+\frac{1}{2}c_8\right)M_{a\pi}^{SP}\Bigr ] \Biggr \}, \label{eq:pi0pi0}
\eeq
\beq
\calm_{\rm LO}(B^+\to \pi^+ \pi^0)&=& \frac{1}{\sqrt{2}}\frac{G_F}{\sqrt{2}}
\Biggl\{\lambda_u\Big[\left(a_1+a_2\right)f_{\pi}F_{e\pi}^{V-A}
+\left(c_1+c_2\right)M_{e\pi}^{V-A}]\nonumber\\
&&\hspace{-1cm}-\lambda_t\Big[\left(\frac{3}{2}a_9+\frac{3}{2}a_{10}\right)f_{\pi}F_{e\pi}^{V-A}
+\left(\frac{3}{2}a_7\right)f_{\pi}F_{e\pi}^{V+A}
+\left(\frac{3}{2}a_8\right)f_{\pi}F_{e\pi}^{SP}\non
&&\hspace{-1cm}+\left(\frac{3}{2}c_9+\frac{3}{2}c_{10}\right)M_{e\pi}^{V-A}
+\left(\frac{3}{2}c_7\right)M_{e\pi}^{V+A}
+\left(\frac{3}{2}c_8\right)M_{e\pi}^{SP}\Big] \Biggr \}, \label{eq:pizpi0}
\eeq
where $\lambda_u = V_{ub}^*V_{ud}$, $\lambda_t =V_{tb}^*V_{td}$, the  Wilson coefficients $a_i$ are the
same as those defined in Ref.~\cite{nlo05}. The eleven decay amplitudes $ F^{V\pm A}_{e\pi,a\pi}, F^{SP}_{e\pi,a\pi}$,
$M_{e\pi,a\pi}^{V\pm A}$ and $M_{a\pi}^{SP}$ in Eqs.~(\ref{eq:pipi1}-\ref{eq:pizpi0})
are obtained by evaluating analytically the Feynman diagrams as shown in Fig.~1 and have been
given for example in Refs.~\cite{lu2001,nlo05}.

In the framework of the pQCD factorization approach, the NLO contributions
should include the following pieces from rather different sources:
\begin{enumerate}
\item[(1)]
The Wilson coefficients $C_i(\mw)$ at NLO level \cite{buras96},
the renormalization group (RG) evolution matrix $U(\mu,\mw,\alpha)$ at NLO level
\cite{buras96} and the strong coupling constant $\alpha_s(\mu)$ at two-loop level
\cite{pdg2012}.

\item[(2)]
The NLO contributions from the vertex corrections (VC), the quark-loops (QL),
and the chromo-magnetic penguin operator $O_{8g}$ (MP) as given in Refs.~\cite{nlo05,npb675,o8g2003,prd85-074004,cheng14b}.

\item[(3)] The NLO twist-2 and twist-3 contributions to the form factors (FF)
of the $B \to \pi$ transition as calculated in Refs.~\cite{prd85-074004,cheng14b}.
\end{enumerate}

The still missing NLO parts in the pQCD approach are the $O(\alpha_s^2)$
contributions from hard spectator diagrams and annihilation diagrams,
as illustrated by the Fig.~5 of Ref.~\cite{fan2013}.
According to the general arguments as presented in Ref.~\cite{nlo05} and
explicit numerical comparisons of the contributions from different sources
for $B \to K \etap$ decays \cite{fan2013}, one generally believe that
these still missing NLO parts are high order corrections to small quantities,
and therefore  could be neglected safely.

For the details of the calculations about those NLO contributions from
the vertex corrections, the quark-loops and the chromo-magnetic Penguins
$O_{8g}$ and the explicit expressions of these NLO contributions, one can see Refs.~\cite{nlo05,o8g2003}.
The NLO vertex corrections can be taken into account by the proper
replacements of the Wilson coefficients $a_i(\mu)$, as presented explicitly for example in Eqs.~(50,51) of Ref.~\cite{fan2013}.
For the NLO contributions from the quark-loops, for example, the corresponding decay amplitudes are of the form
\beq
\calm^{\rm (QL)}(B^0\to \pi^0\pi^0)&=&
\frac{G_F}{\sqrt{2}}\frac{8\pi}{\sqrt{6}}C_f^2M_B^4\int^1_0dx_1dx_2dx_3\int^\infty_0 b_1db_1b_3db_3\phi_B(x_1)\non
&& \hspace{-2.5cm}\times \Bigl \{ \Bigl [ (1+x_3)\phi_{\pi}^A(x_2)\phi_{\pi}^A(x_3)
+r_\pi(1-2x_3) \left ( \phi_{\pi}^P(x_3)\phi_{\pi}^A(x_2)+\phi_{\pi}^T(x_3)\phi_{\pi}^A(x_2) \right )\non
&& \hspace{-2.5cm}+2r_\pi\phi_{\pi}^A(x_3)\phi_{\pi}^P(x_2)\Bigr ] \cdot  E^{ql}(t_q,l^2) \cdot h_e(x_1,x_3,b_1,b_3) \non
&&\hspace{-2.5cm} + \Bigl [ 2r_\pi x_1 \phi_{\pi}^A(x_3) \phi_{\pi}^P(x_2) +2r_\pi \phi_{\pi}^P(x_3) \phi_{\pi}^A(x_2) \Bigr ]
\cdot E^{ql}(t_q^\prime,l^2) \cdot h_e(x_3,x_1,b_3,b_1) \Bigr \},
\label{eq:m001}
\eeq
\beq
\calm^{\rm (QL)}(B^0\to \pi^+\pi^-)&=&\frac{G_F}{\sqrt{2}}\frac{8\pi}{\sqrt{6}}C_f^2M_B^4\int^1_0dx_1dx_2dx_3\int^\infty_0 b_1db_1b_3db_3\phi_B(x_1)\non
&&\hspace{-2cm}\times \Bigl \{\Bigl [(1+x_3)\phi_{\pi}^A(x_2)\phi_{\pi}^A(x_3)
+r_\pi (1-2x_3) \left (\phi_{\pi}^P(x_3)\phi_{\pi}^A(x_2)+\phi_{\pi}^T(x_3)\phi_{\pi}^A(x_2) \right )\non
&&\hspace{-2cm} +2r_\pi\phi_{\pi}^A(x_3)\phi_{\pi}^P(x_2)\Bigr ]\cdot E^{ql}(t_q,l^2)\cdot h_e(x_1,x_3,b_1,b_3)\non
&&\hspace{-2cm}+\left[2r_\pi x_1 \phi_{\pi}^A(x_3) \phi_{\pi}^P(x_2)
+2r_\pi \phi_{\pi}^P(x_3) \phi_{\pi}^A(x_2) \right]\cdot E^{ql}(t_q^\prime,l_2)h_e(x_3,x_1,b_3,b_1)\Bigr\},
\\
\calm^{\rm (QL)}(B^+\to \pi^+\pi^0)&=&0,
\eeq
where $r_\pi=m_0^{\pi}/m_B$, and the terms proportional to $r_\pi^2$ are not shown in above equations.
The function $E^{ql}(t_q,l^2), h_e(x_i,b_i)$ and other relevant parameters can be found
for example in Appendix B of Ref.~\cite{fan2013}. It is straightforward to find the
NLO contributions $\calm^{(MP)}(B\to \pi\pi)$ from the $O_{8g}$ insertion correction \cite{o8g2003,nlo05,fan2013}.

Very recently, the NLO twist-2 and twist-3 contributions to the form factors
$f^{+,0}(q^2)$ of $B \to \pi$ transition have been calculated in Refs.~\cite{prd85-074004,cheng14b}.
When these NLO contributions are taken into account, the form factor
$f^{+}(q^2)$, for example, can be written in the form of
\beq
f^+(q^2)|_{\rm NLO} &=& 8 \pi m^2_B C_F \int{dx_1 dx_2} \int{b_1 db_1 b_2 db_2} \phi_B(x_1,b_1)\non
&& \hspace{-2cm}\times \Biggl \{ r_\pi
\left [\phi_{\pi}^{P}(x_2) - \phi_{\pi}^{T}(x_2) \right ]
\cdot \alpha_s(t_1)\cdot e^{-S_{B\pi}(t_1)}\cdot S_t(x_2)\cdot h(x_1,x_2,b_1,b_2) \non
&&\hspace{-2cm}  + \Bigl [ (1 + x_2 \eta)
\left (1 + F^{(1)}_{\rm T2}(x_i,\mu,\mu_f,q^2)\; \right )
\phi_{\pi}^A(x_2)+ 2 r_\pi \left ( \frac{1}{\eta} - x_2 \right )\phi_{\pi}^T(x_2) - 2x_2 r_\pi \phi_{\pi}^P(x_2) \Bigr ]
\non
&&
\hspace{-1.5cm} \cdot \alpha_s(t_1)\cdot e^{-S_{B\pi}(t_1)} \cdot S_t(x_2)\cdot h(x_1,x_2,b_1,b_2)\non
&& \hspace{-2cm} + 2 r_{\pi} \phi_{\pi}^P(x_2)
\left (1 + F^{(1)}_{\rm T3}(x_i,\mu,\mu_f,q^2)\right ) \cdot \alpha_s(t_2)\cdot e^{-S_{B\pi}(t_2)}
\cdot S_t(x_2)\cdot h(x_2,x_1,b_2,b_1) \Biggr \},
\label{eq:ffnlop}
\eeq
with the NLO twist-2 and twist-3 correction factors
\beq
F^{(1)}_{\rm T2}(x_i,\mu,\mu_f,q^2)&=& \frac{\alpha_s(\mu_f) C_F}{4 \pi}
\Biggl [\frac{21}{4} \ln{\frac{\mu^2}{m^2_B}}
-(\frac{13}{2} + \ln{r_1}) \ln{\frac{\mu^2_f}{m^2_B}}
+\frac{7}{16} \ln^2{(x_1 x_2)}+ \frac{1}{8} \ln^2{x_1} \non
&&+ \frac{1}{4} \ln{x_1} \ln{x_2}
+ \left (- \frac{1}{4}+ 2 \ln{r_1} + \frac{7}{8} \ln{\eta} \right ) \ln{x_1}
+ \left (- \frac{3}{2} + \frac{7}{8} \ln{\eta} \right) \ln{x_2} \non
&&+ \frac{15}{4} \ln{\eta} - \frac{7}{16} \ln^{2}{\eta}
+ \frac{3}{2} \ln^2{r_1} - \ln{r_1}
+ \frac{101 \pi^2}{48} + \frac{219}{16} \Biggr ],  \label{eq:ffnlot2}\\
F^{(1)}_{\rm T3}(x_i,\mu,\mu_f,q^2)&=&\frac{\alpha_s(\mu_f) C_F}{4 \pi}
\Biggl [\frac{21}{4} \ln{\frac{\mu^2}{m^2_B}}
- \frac{1}{2}(6 + \ln{r_1}) \ln{\frac{\mu^2_f}{m^2_B}}
+ \frac{7}{16} \ln^2{x_1} - \frac{3}{8} \ln^2{x_2} \non
&& \hspace{-1cm}+ \frac{9}{8} \ln{x_1} \ln{x_2}
+ \left (- \frac{29}{8}+ \ln{r_1} + \frac{15}{8} \ln{\eta} \right ) \ln{x_1}
+ \left (- \frac{25}{16} + \ln{r_2} + \frac{9}{8} \ln{\eta} \right) \ln{x_2} \non
&&\hspace{-1cm}+ \frac{1}{2} \ln{r_1} - \frac{1}{4} \ln^{2}{r_1} + \ln{r_2}
- \frac{9}{8} \ln{\eta} - \frac{1}{8} \ln^{2}{\eta} + \frac{37 \pi^2}{32}
+ \frac{91}{32} \Biggr ],
\eeq
where $r_i=m_B^2/\xi_i^2$ with the choice of $\xi_1=25 m_B$ and $\xi_2=m_B$\cite{prd85-074004},
$\eta=1-q^2/m_B^2$  with $q^2=(P_1-P_3)^2$ is the energy fraction carried by
the meson which picks up the spectator quark of $B$ meson, $\mu$ ($\mu_f$) is  the renormalization (factorization ) scale,
the hard scale $t_{1,2}$ are chosen as the largest scale of the propagators in the hard $b$-quark decay
diagrams \cite{prd85-074004,cheng14b}, the function $S_t(x_2)$ and the hard function
$h(x_i,b_j)$ can be found in Refs.~\cite{prd85-074004,cheng14b}.
For $B \to \pi \pi$ decays, the large recoil region
corresponds to the energy fraction  $\eta  \sim \textit{O}(1)$. We here also set
$\mu=\mu_f=t$ in order to minimize the NLO contribution to the form factors
\cite{prd83-054029,cheng14b}.

In the numerical calculations, we use the following input parameters
\cite{hfag2012,pdg2012} (all masses and decay constants in units of GeV)
\beq
f_B &=& 0.21, \quad f_\pi= 0.13, \quad m_\pi=0.14, \quad m_{0\pi} = 1.4,\quad
M_B = 5.28,\quad  m_b = 4.8, \non
m_c &=& 1.5, M_W = 80.41, \quad \tau_{B^0} = 1.53 {\rm ps},
\quad \tau_{B^+} = 1.641 {\rm ps}. \label{eq:para}
\eeq
For the CKM matrix elements, we adopt the Wolfenstein parametrization
with the CKM parameters as given in Ref.~\cite{pdg2012}:
$ A = 0.832 \pm 0.017$, $\lambda = 0.2246 \pm 0.0011$,
$\bar{\rho} = 0.130 \pm 0.018$ and $\bar{\eta} = 0.350 \pm 0.013$.

We firstly calculate the pQCD predictions for the form factor  $F_0^{B \to \pi}(0)$ for $B \to \pi$ transition at the
LO and NLO level respectively and find numerically that
\beq
F_0^{B \to \pi}(0)=\left\{ \begin{array}{ll}
0.27\pm 0.05, & {\rm LO}, \\ 0.28^{+0.05}_{-0.06}, & {\rm NLO.} \\ \end{array} \right.
\label{eq:f0bpi}
\eeq
We find that the NLO twist-2 and twist-3 contribution  are similar in magnitude but have opposite sign,
the $\sim 15\%$ enhancement to the central value of the LO pQCD prediction is therefore largely canceled by the
inclusion of the NLO twist-3 contribution. The pQCD predictions as given in Eq.~(\ref{eq:f0bpi})
agree very well with those obtained from the QCD sum rule or other methods.

Using the input parameters  and the wave functions as given in previous sections,
it is easy to calculate the CP-averaged branching ratios for
the considered three $B \to \pi \pi$ decays. When all currently known
NLO contributions are taken into account, we find the following NLO pQCD
predictions for the CP-averaged branching ratios:
\beq
Br( B^0 \to \pi^+\pi^-) &=& \left [7.67^{+2.63}_{-1.85}(\omega_b)^{+1.53}_{-1.39}
(f_B)^{+1.44}_{-1.28}(a^{\pi}_2) \right ]\times 10^{-6} ,\non
Br( B^+ \to \pi^+\pi^0) &=& \left [4.27^{+1.42}_{-1.01}(\omega_b)^{+0.85}_{-0.77}
(f_B)^{+0.82}_{-0.75}(a^{\pi}_2) \right ]\times 10^{-6} , \non
Br( B^0 \to \pi^0\pi^0) &=& \left [0.23^{+0.08}_{-0.05}(\omega_b)^{+0.05}_{-0.04}
(f_B)^{+0.04}_{-0.03}(a^{\pi}_2) \right ]\times 10^{-6} ,
\label{eq:br-2pi1}
\eeq
where the major theoretical errors are induced by the
uncertainties of $\omega_b=0.4 \pm 0.04$ GeV, $f_B=0.21\pm 0.02$GeV and
Gegenbauer moment $a^{\pi }_2=0.25\pm 0.15$, respectively.

In Table ~\ref{br1}, we show the pQCD predictions for the CP-averaged
branching ratios of the three
$B \to \pi\pi$ decays when the NLO contributions from different sources
are included step by step.
The label ``NLOWC" means the pQCD predictions from the LO Feynman diagrams as illustrated
in Fig.~1 but calculated numerically by using the Wilson coefficients $C_i(\mw)$
and the RG evolution matrix $U(t,m,\alpha)$ at the NLO level.
The label ``+VC", ``+QL" and ``+MP" means the ``NLOWC" results plus the
NLO contribution from the vertex corrections(VC), the quark loops(QL) and
the chromo-magnetic penguin(MP), respectively.
The label ``NLO" means all currently known NLO contributions, including the very
recently known NLO twist-2 and twist-3 contributions to the $B \to \pi$ transition
form factor \cite{prd85-074004,cheng14b}, are all taken into account and all theoretical errors
from different sources are added in quadrature.
In the last two columns of Table \ref{br1}, for the sake of comparison, we also list
the measured values as given by HFAG \cite{hfag2012} and those QCDF predictions as
given in Ref.~\cite{npb675}.

\begin{table}[thb]
\begin{center}
\caption{ The pQCD predictions for the CP-averaged branching ratios (in unit of $10^{-6}$).
The meaning of the labels have been explained in the text.}
\label{br1} \vspace{0.2cm}
\begin{tabular}{l|c c c c c c| c l} \hline \hline
Channel & LO &NLOWC& +VC & +QL & +MP &  NLO & QCDF\cite{npb675} & Data\cite{hfag2012} \\ \hline
$B^0 \to \pi^+\pi^-$  &$7.46$ &$6.65$&$6.91$ &$7.02$ &$6.87$ &$7.67^{+3.47}_{-2.64}$ &$8.9$&$5.10\pm 0.19$\\
$B^+ \to \pi^+\pi^0$  &$3.54$ &$4.23$&$3.54$ &$-$   &$-$   &$4.27^{+1.85}_{-1.47}$ &$6.0$  &$5.48^{+0.35}_{-0.34}$  \\
$B^0 \to \pi^0\pi^0$  &$0.12$ &$0.24$&$0.27$ &$0.29$ &$0.21$ &$0.23^{+0.19}_{-0.15}$ &$0.3$&$1.91^{+0.22}_{-0.23}$\\
\hline\hline
\end{tabular}
\end{center} \end{table}

Now we turn to the evaluations of the CP-violating asymmetries of $B
\to  \pi \pi$ decays in pQCD approach. For $B^+ \to \pi^+ \pi^0$ decays,
the LO and NLO pQCD predictions for the direct CP-violating asymmetries
$\acp$ are the following
\beq
\acp^{dir}(B^\pm\to \pi^\pm \pi^0)= \left\{ \begin{array}{cc}
-4.7\%, & {\rm LO}, \\ -5.6\%, & {\rm NLO} \\ \end{array}. \right.
\eeq

\begin{table}[thb]
\begin{center}
\caption{ The LO and NLO pQCD predictions for the direct and mixing-induced
CP asymmetries for $B^0 \to \pi^+\pi^-$  and $\pi^0\pi^0$
decays. The world averages as given in Ref.~\cite{hfag2012} are listed in last column.}
\label{cp2}
\vspace{0.2cm}
\begin{tabular}{l |c c c c c c| c} \hline \hline
Mode&  LO &NLOWC & +VC & +QL & +MP & NLO & Data \cite{hfag2012} \\
\hline
${\cal A}_{\pi\pi}$ &$0.27$&$0.25$ &$0.26$ &$0.13$ &$0.12 $ &$0.12^{+0.04}_{-0.06}$& $+0.31\pm 0.05$\\
${\cal S}_{\pi\pi}$ &$-0.28$&$-0.40$ &$-0.39$ &$-0.49$ &$-0.41$ &$-0.40^{+0.05}_{-0.04}$& $-0.66\pm 0.06$\\
\hline
$\acp^{dir}(\pi^0\pi^0)$ &$-0.10$ &$-0.51$ &$0.61$ &$0.69 $ &$0.74$  &$ 0.78^{+0.05}_{-0.08}$& $-$ \\
$\acp^{mix}(\pi^0\pi^0)$ &$-0.02$ &$0.61$ &$0.67$ &$0.41$ &$0.50$ &$0.47^{+0.02}_{-0.11}$& $-$\\
\hline
\end{tabular}\end{center}
\end{table}


For $B^0 \to \pi^+\pi^-$ and $\pi^0\pi^0$ decays, the time-dependent decay rate
is defined as \cite{belle2013}
\beq
{\cal P}(\Delta t, q) = \frac{ e^{-|\Delta t|/\tau_{B^0} } }{ 4\tau_{B^0} }
\biggl \{ 1  + q \bigl[ {\cal A}_{\pi\pi} \; \cos (\Delta m_d \Delta t)
+ {\cal S}_{\pi\pi} \sin (\Delta m_d \Delta t )\bigr]\biggr\}.
\label{eq:acp2}
\eeq
where $\Delta t = t_{\pi\pi} - t_{\rm tag}$, $\tau_{B^0}$ is the $B^0$ lifetime,
$\Delta m_d$ is the mass difference between the two mass eigenstates of the
neutral $B^0$ meson, and $q=+1(-1)$ when $f_{\rm tag } = B^0(\bar{B}^0 ) $.
The parameter ${\cal A}_{\pi\pi} $ and ${\cal S}_{\pi\pi}$
are the direct and mixing-induced $CP$-violating parameters respectively, and have been
defined as the form of
\beq
{\cal A}_{\pi\pi}=\frac{|\lambda_{\pi\pi}|^2-1}{1+ |\lambda_{\pi\pi}|^2 },
\quad {\cal S}_{\pi\pi}= \frac{2 {\rm Im}(\lambda_{\pi\pi})}{1+ |\lambda_{\pi\pi}|^2 },
\eeq
where $\lambda_f =\frac{q}{p}\frac{\ov{A}_f}{A_f}$ depends on the parameters
related to the $B^0-\bar{B}^0$ mixing and to the decay amplitudes of $B^0/\bar{B}^0 \to f$
with the CP eigenstate  $f$.

Using the input parameters and the wave functions as given in previous sections,
we calculate the CP-violating asymmetries for $B^0\to (\pi^+\pi^-,\pi^0\pi^0)$
decays and list the numerical results in Table ~\ref{cp2}.
The labels ``NLOWC", ``+VC", ``+QL" , ``+MP" and ``NLO" in Table ~\ref{cp2}
have the same meaning as those in Table ~\ref{br1}.
The major theoretical errors as given in Table ~\ref{cp2}
are induced by the uncertainties of input parameters of $\omega_b$, and $a_2^\pi $.
As a comparison, we also list currently available measured values
for ${\cal A}_{\pi\pi}$ and ${\cal S}_{\pi\pi}$ for $B^0\to \pi^+\pi^-$ decay in last column.

From the numerical values as listed in Table ~\ref{br1} and \ref{cp2}, one can
see the following points:
\begin{enumerate}
\item[(i)]
For the decay rates $Br( B^0 \to \pi^+\pi^-)$ and
$Br( B^0 \to \pi^+\pi^-)$, the NLO pQCD predictions agree with the data within $1\sigma$ error
since the theoretical errors are still large.

\item[(ii)]
For $B^0 \to \pi^0 \pi^0$ decay, although the NLO contributions provide
about $\sim 100\%$ enhancement to the LO result, it is still much smaller than
the measured one. The so-called "$\pi\pi$" puzzle is still an open problem.
The contribution from the soft Glauber gluon\cite{li2011}, or the inclusion
of  the charm content effect through the tetramixing of $\pi$-$\eta$-$\eta'$-$\eta_c$
as proposed in Ref.~\cite{Peng:2011ue},
may be the possible ways out of this crisis, but it needs more studies.

\item[(iii)]
For CP-violating asymmetries of $B^0\to \pi^+\pi^-$ decay, the pQCD predictions for ${\cal A}_{\pi\pi}$ and
${\cal S}_{\pi\pi}$ agree with the measured values in both the sign and magnitude,
but have a little smaller central values.

\item[(iv)]
For $B^+\to \pi^+\pi^0$ decay, its direct CP violation is small in size.
For $B^0\to \pi^0\pi^0$ decay, however, the pQCD predictions for their CP-violating asymmetries are large in size
and may be measurable in the running LHCb and future super-B experiments.

\end{enumerate}

From the numerical results as listed in Table I-II, one can see that the
LO pQCD predictions could be changed significantly after the inclusion of the NLO contributions.
We  here will check the relative strength for those LO contributions from different kinds of Feynman diagrams, and then
examine  the  effects of the NLO contributions from different sources.

\begin{table}[thb]
\begin{center}
\caption{The LO pQCD predictions for the numerical values (in unit of $10^{-4}$)
of the individual and total decay amplitudes of $B^0/\ov{B}^0
\to (\pi^+\pi^-,\pi^0\pi^0)$  and $B^\pm\to \pi^\pm\pi^0$ decays, as well as the
ratios $R_{\rm LO}$.} \label{n01-lo}
\vspace{0.2cm}
\begin{tabular}{l c c c c |c } \hline \hline
Decay& $\calm^{\rm a+b}$&$\calm^{\rm c+d}$& $\calm^{\rm anni}$&$\calm_{\rm LO}$&$R_{\rm LO}$\\ \hline
$B^0\to  \pi^+\pi^-$ & $-1.40 -i 2.32$  & $0.094 +i 0.022$&$0.11 +i 0.48$&$-1.19 -i 1.81$ & $7.33:0.009:0.25:4.72$ \\
$B^+\to  \pi^+\pi^0$ & $-0.61 -i 1.50$  & $-0.073 -i 0.048$ &$-$ &$-0.69 -i 1.54$& $2.62:0.008:0.00:2.85$ \\
$B^0\to  \pi^0\pi^0$ & $-0.31 -i 0.05$  & $0.13 +i 0.08$&$0.01 +i 0.26$&$-0.17 +i 0.29$& $0.10:0.020:0.07:0.11$ \\
\hline\hline
$\ov{B}^0\to  \pi^+\pi^-$ & $-1.40 +i 2.32$  & $-0.041 -i 0.090$&$0.11 +i 0.30$&$-1.33 +i 2.53$ & $7.33:0.010:0.10:8.16$ \\
$B^-\to  \pi^-\pi^0$      & $-0.61 +i 1.50$  & $0.020  +i 0.085$&$-$           &$-0.59 +i 1.58$ & $2.62:0.008:0.00:2.85$ \\
$\ov{B}^0\to  \pi^0\pi^0$ & $-0.31 +i 0.05$  & $-0.03 -i 0.15 $&$0.05 +i 0.19$&$-0.29 +i 0.08$ & $0.10:0.020:0.04:0.30$ \\
\hline\hline
\end{tabular}
\end{center} \end{table}

In Table  \ref{n01-lo} we show the central values of the
pQCD predictions for the numerical values
(in unit of $10^{-4}$) of the decay amplitude from different Feynman
diagrams at the LO level.
The label ``$\calm^{a+b}$"  ( ``$\calm^{c+d}$" ) means the decay amplitude
of the factorizable emission diagrams Fig.1(a) and 1(b) ( the spectator diagrams
Fig.1(c) and 1(d) ).
The label ``$\calm^{anni}$" means the decay amplitude
from the four annihilation diagrams Fig.1(e) - 1(h). The term $\calm_{\rm LO}$
means the full LO decay amplitude. The ratio $R_{\rm LO}$ in Table  \ref{n01-lo} is
defined as the form of
\beq
R_{\rm LO}&=& |\calm^{a+b}|^2: |\calm^{c+d}|^2:|\calm^{anni}|^2:|\calm_{\rm LO}|^2.
\label{eq:r1-lo}
\eeq
From the numerical results as listed in Table \ref{n01-lo}, one can find the following points:

\begin{enumerate}
\item[(i)]
At the leading order, the two factorizable emission diagrams
do provide the dominant contribution.
For $B^0/\bar{B}^0\to \pi^+\pi^-$ and $B^\pm \to \pi^\pm \pi^0$ decays,
we find numerically that
\beq
|\calm^{a+b}|^2 \gg  |\calm^{c+d}|^2 \quad {\rm or} \quad |\calm^{anni}|^2.
\eeq
For $B^0/\bar{B}^0\to \pi^0 \pi^0$ decay, although $|\calm^{a+b}|^2$ is still larger
than $|\calm^{anni}|^2$, the annihilation diagrams for this decay do have
a small real part but a large imaginary part, which in turn result in an
effective contribution to its branching ratio and also provide the large strong phase
required to produce the large CP violation.

\item[(ii)]
By comparing $\calm^i$ for $B^0/B^+$ decays and their CP conjugated $\bar{B}^0/B^-$ decays,
one can see that the amplitude $\calm^{a+b}$ does not has the strong phase,
$\calm^{c+d}$ has a small strong phase,  but the annihilation diagrams (i.e.,
$\calm^{anni}$ ) do provide the dominant large strong phase. This
feature confirmed the general expectation again \cite{fan2013} in the pQCD
factorization approach: The strong phase needed to produce large CP violation for the two-body charmless hadronic
B meson decays really comes from the annihilation diagrams.

\end{enumerate}

In Table \ref{nlo-02}  the label ``$\Delta\calm_{\rm FF}$" describes the total modification due to the
inclusion of both the NLO twist-2 and twist-3 contributions to the $B\to \pi$ transition form factors
\cite{prd85-074004,cheng14b}, it is indeed very small in size due to the strong cancelation between
the NLO twist-2 and twist-3 part. The label ``$\Delta \calm_{\rm NLO}$" denotes the changes with respect to
``$\calm_{\rm LO}$" induced by the inclusion of all currently known NLO contributions,
and finally we define the total decay amplitude at the NLO level as $\calm_{\rm NLO} = \calm_{\rm LO}
+ \Delta \calm_{\rm NLO}$ and the ratio $R_{\rm NLO}$ as $R_{\rm NLO} = |\calm_{\rm NLO}|^2/|\calm_{\rm LO}|^2$,
which measures the effects of the NLO contributions to the considered decays directly.


\begin{table}[]
\begin{center}
\caption{The same as in Table \ref{n01-lo} but for $\Delta \calm_{\rm FF}$,
$\Delta \calm_{\rm NLO}$ and $\calm_{\rm NLO}$ for $B/\ov{B} \to \pi\pi$ decays.
The ratios $R_{\rm NLO}$ are also listed in last column. }
\label{nlo-02}
\vspace{0.2cm}
\begin{tabular}{l|cc c c| c } \hline \hline
Decay &  $\calm_{\rm LO}$ & $\Delta \calm_{\rm FF}$&  $\Delta \calm_{\rm NLO}$& $\calm_{\rm NLO}$ & $R_{\rm NLO}$\\
\hline
$B^0\to \pi^+\pi^-$  &$-1.20 -i 1.82$&$-0.07 -i 0.13$&$-0.17 -i 0.17$& $-1.37 -i 1.99$&$1.23$\\
$B^+\to  \pi^+\pi^0$ &$-0.69 -i 1.54$&$-0.05 -i 0.08$&$-0.24 -i 0.06$& $-0.93 -i 1.60$&$1.20$\\
$B^0\to  \pi^0\pi^0$ &$-0.17 +i 0.29$&$ 0.00 -i 0.01$&$ 0.12 -i 0.09$& $-0.05 +i 0.20$&$0.38$\\
\hline\hline
$\ov{B}^0\to \pi^+\pi^-$ &$-1.33 +i 2.53$&$-0.08 +i 0.12$ &$-0.14 -i 0.23$& $-1.47 +i 2.30$&$0.92$ \\
$B^-\to  \pi^+\pi^0$     &$-0.59 +i 1.58$&$-0.03 +i 0.10$ &$ 0.12 +i 0.21$& $-0.47 +i 1.79$&$1.20$ \\
$\ov{B}^0\to \pi^0\pi^0$ &$-0.29 +i 0.09$&$-0.02 +i 0.01$ &$-0.28 -i 0.26$& $-0.57 -i 0.17$&$3.85$ \\
\hline\hline
\end{tabular}
\end{center} \end{table}

From the pQCD predictions for the numerical values of the decay amplitudes
as listed in Table  \ref{nlo-02}, we find the following points:
\begin{enumerate}
\item[(i)]
As illustrated by the numbers in third column, the contributions from the NLO  contributions to the
$B \to \pi$ transition form factors are indeed very small. The reason id  the large cancelation
between the NLO twist-2 and twist-3 pieces.

\item[(ii)]
For $B^\pm \to \pi^\pm \pi^0$ decays, the inclusion of all NLO contributions leads to a
$20\%$ enhancement to the LO one.
For $B^0/\ov{B}^0 \to \pi^+ \pi^-$ decay, the effects of NLO contribution to
the decay amplitude of the $B^0\to \pi^+ \pi^-$ decay and its CP conjugated
decay are rather different: about $20\%$ enhancement to the former case,
but $9\%$ decrease to $\bar{B}^0 \to \pi^+\pi^-$ decay mode. And
finally provide a $3\%$ enhancement to its CP-averaged branching ratio.

\item[(iii)]
For $B^0/\ov{B}^0 \to \pi^0 \pi^0$ decays, the NLO contributions themselves and
their effects on the LO decay amplitudes are rather different for $B^0 \to \pi^0 \pi^0$ decay
and its CP-conjugated decay mode:
\beq
\Delta \calm_{\rm NLO}&=&  \left\{ \begin{array}{cc}
0.12-i 0.09, & {\rm for} \ \  B^0\to \pi^0\pi^0, \\
-0.28 -i0.26, & {\rm for} \ \ \ov{B}^0\to \pi^0\pi^0, \\ \end{array}. \right.\\
R_{\rm NLO}&=&  \left\{ \begin{array}{cc}
0.38, & {\rm for} \ \  B^0\to \pi^0\pi^0, \\
3.85, & {\rm for} \ \ \ov{B}^0\to \pi^0\pi^0, \\ \end{array}. \right.
\eeq
due to the very different interference patterns between $\calm_{\rm LO}$ and
$\Delta \calm_{\rm NLO}$ for these two decay modes. The total enhancement to the CP-averaged
decay rate $Br(B^0/\ov{B}^0 \to \pi^0 \pi^0 )$ is around $100\%$.

\end{enumerate}

In short, we made a systematic study for the $B \to  \pi \pi$
decays in the pQCD factorization approach with the inclusion of all
currently known NLO contributions to the considered decays. We find the following points
\begin{enumerate}
\item[(i)]
For $B^0 \to \pi^+\pi^-$ and $\pi^+\pi^0$ decays, the NLO pQCD predictions for their
CP-averaged branching ratios and CP violating asymmetries agree well with the measured values
within one standard deviation.

\item[(ii)]
For the CP-averaged branching ratio $Br(B^0/\ov{B}^0 \to \pi^0 \pi^0)$, however, although the NLO contributions
can provide a $\sim 100\%$ enhancement to the LO result, it is still much smaller than
the measured one. The so-called "$\pi\pi$" puzzle is still an open problem.

\item[(iii)]
We examined the relative strength for those LO and NLO contributions from different sources.

\end{enumerate}

\begin{acknowledgments}

This work is supported by the National Natural Science Foundation of China under the Grant No.~11235005.

\end{acknowledgments}


\end{document}